\documentclass[journal=jacsat,manuscript=article,email=false]{achemso}

\makeatletter
\renewcommand*{\acs@author@fnsymbol@symbol}[1]{%
\ifcase #1 *\or
    {1}\or
    {2}\or
    {3}\or
        \ast\or
        \dagger\or
            \ddagger\or
            \bot\or
                \#\or
                @\or
                    \triangle\or
                    \nabla
    \fi
}
\usepackage[version=3]{mhchem} 
\mciteErrorOnUnknownfalse 

\newcommand{\physicsDept}{ Department of Physics, Laboratory of Atomic and Solid State Physics, Cornell University, Ithaca, New York 14853, USA}

\newcommand{\mseDept}{ Department of Materials Science and Engineering, Cornell University, Ithaca, NY 14853, USA}

\newcommand{\kavli}{ Kavli Institute at Cornell for Nanoscale Science, Ithaca, NY 14853, USA}

\newcommand{\Uchicago}{ Current Address: Pritzker School of Molecular Engineering, The University of Chicago, Chicago, IL 60637, USA}

\newcommand{\FeSeSTO}{FeSe/SrTiO$_3$ }
\newcommand{\FeSeSTOns}{FeSe/SrTiO$_3$}

\newcommand{\Tc}{$T_{c}$ }
\newcommand{\TCnospace}{$T_c$}

\newcommand{\Tgap}{$T_{\Delta}$ }
\newcommand{\Tgapns}{$T_{\Delta}$}
\newcommand{\Tonset}{$T_{\mathrm{onset}}$ }
\newcommand{\Tonsetns}{$T_{\mathrm{onset}}$}
\newcommand{\Tzero}{$T_{0}$ }
\newcommand{\Tzerons}{$T_{0}$}
\newcommand{\Tbkt}{$T_{BKT}$ }
\newcommand{\Tbktns}{$T_{BKT}$}
\newcommand{\Tstar}{$T^{\ast}$ }
\newcommand{\Tstarns}{$T^{\ast}$}
\newcommand{\insitu}{\emph{in situ} }

\newcommand{\exsitu}{\emph{ex situ} }
\newcommand{\Rsheet}{$R_{s}$ }

\author{B.D. Faeth}
\affiliation{\physicsDept}
\author{S.-L. Yang}\altaffiliation{\Uchicago} 
\affiliation{\physicsDept}
\alsoaffiliation{\kavli}
\alsoaffiliation{\mseDept}
\author{J.K. Kawasaki}
\affiliation{\physicsDept}
\alsoaffiliation{\kavli}
\altaffiliation{ Current Address: Department of Materials Science and Engineering, University of Wisconsin, Madison, Wisconsin 53706, USA.} 
\author{J.N. Nelson}
\affiliation{\physicsDept}
\author{P. Mishra}\altaffiliation{ Current Address: Department of Physics, Indian Institute of Science, Bangalore 560012, India}\affiliation{\physicsDept}
\author{C. Li}
\affiliation{\physicsDept}
\author{D.G. Schlom}
\affiliation{\mseDept}
\author{K.M. Shen}
\affiliation{\physicsDept}
\alsoaffiliation{\kavli}

\title{Incoherent Cooper pairing and pseudogap behavior in single-layer \FeSeSTOns}

\begin{document}

\textbf{In many unconventional superconductors, the presence of a pseudogap - a suppression in the electronic density of states extending above the critical temperature - has been a long-standing mystery. Here, we employ combined \textit{in situ} electrical transport and angle-resolved photoemission spectroscopy (ARPES) measurements to reveal an unprecedentedly large pseudogap regime in single-layer \FeSeSTOns, an interfacial superconductor where incoherent Cooper pairs are initially formed above \Tgap $\approx$ 60 K, but where a zero resistance state is only achieved below \Tzero $<$ 30 K. We show that this behavior is accompanied by distinct transport signatures of two-dimensional phase fluctuating superconductivity, suggesting a mixed vortex state hosting incoherent Cooper pairs which persist well above the maximum clean limit \Tc of $\approx$ 40 K. Our work establishes the critical role of reduced dimensionality in driving the complex interplay between Cooper pairing and phase coherence in two-dimensional high-$T_c$ superconductors, providing a paradigm for understanding and engineering higher-\Tc interfacial superconductors.}

Single-layer FeSe grown on SrTiO$_{3}$ (\FeSeSTOns) has attracted interest due to its characteristics as an atomically thin, interfacially enhanced high-\Tc superconductor~\cite{Wang2012}. \FeSeSTO exhibits a spectroscopic gap opening temperature (\Tgapns) between 60 to 70 K~\cite{Liu_origin_2012,Yang2019,Lee2014,Song2019}, nearly one order of magnitude higher than that of bulk FeSe (8K)~\cite{Hsu14262}, and in excess of related electron-doped FeSe-based bulk compounds ($\approx$ 40 K)~\cite{Zhao2016,Zhang2011}. The combination of its high \TCnospace, relative simplicity, and inherently two-dimensional (2D) nature positions \FeSeSTO as an ideal platform for exploring the importance of superconducting fluctuations and the possibility of interfacial enhancement in high-$T_c$ materials.

Nevertheless, significant challenges impede the systematic study of \FeSeSTOns, as its air sensitivity, variability in the post-growth annealing process, and potential impact of capping layers make meaningful comparisons across different techniques and studies, both \insitu and \exsitu, difficult~\cite{Xing_2015,He2013}.  Consequently, there remains a widely observed but heretofore unexplained discrepancy between the gap opening temperature $T_{\Delta}$ observed by ARPES ($T_{\Delta}$ $\approx$ 60 K) and the temperature at which a zero-resistance state has been measured by electrical transport, \Tzero (\Tzero $<$ 30 K)~\cite{Zhang2014,Yang2019,Zhao_2016,Zhao_SciAdv_2018,Wang_2017}. A potential resolution to this puzzle is the existence of Cooper pair fluctuations above $T_c$ which are known to play an important role in two-dimensional superconductors as well as underdoped cuprates, but have not been widely investigated for \FeSeSTOns.

To reveal the intrinsic nature of superconductivity and the pseudogap in \FeSeSTOns, we employ for the first time, a combination of angle-resolved photoemission spectroscopy (ARPES) and \insitu resistivity measurements to simultaneously probe both the spectroscopic and electrical transport properties of pristine single-layer \FeSeSTO samples in ultrahigh vacuum.  Through a systematic investigation of a large number of such samples, we reveal the presence of intrinsic superconducting fluctuations over an unprecedentedly broad temperature range, as characterized by the window between the onset of spectroscopic gap \Tgap and the onset of zero resistance \Tzerons.  This result establishes the essential role that reduced dimensionality plays in the superconductivity of \FeSeSTO and resolves the long-standing confusion surrounding the critical temperature of \FeSeSTOns.

In Fig. 1, we show combined \insitu resistivity and ARPES measurements conducted on the same sample of single-layer \FeSeSTOns. The Fermi surface (Fig. 1a) is comprised of electron pockets centered at the $M$ point consistent with an electron doping of $0.11 e^{-}$ per unit cell, in good agreement with earlier reports~\cite{Liu_origin_2012,Tan2013,Lee2014}, and exhibits the expected spectroscopic signatures of superconductivity (a well-defined gap and band back-bending).  In Fig. 1c, we show the sheet resistance, $R_{s}(T)$ which exhibits a hump-like feature at 280 K, characteristic of heavily electron-doped bulk FeSe-derived compounds~\cite{Sun2018}, and a broad superconducting transition which onsets at \Tonset = 44$\pm$3 K, eventually falling below 0.1\% of $R_{70\mathrm{K}}$  at \Tzero = 29$\pm$0.2 K.  When measured \emph{in situ}, \FeSeSTO samples exhibit residual resistivity ratios (RRRs, defined as $R_{300 \mathrm{K}}/R_{70 \mathrm{K}}$) of $\approx$10, in contrast to RRRs of $\approx$ 1-2 for capped single-layer films reported in the literature~\cite{Zhang2014}. While samples remain robust for hundreds of hours and over numerous cooling/warming cycles when maintained under ultrahigh vacuum (red curve, Fig. 1d), pristine films deteriorate instantaneously upon exposure to atmosphere (black curve, Fig. 1d).

To explore this behavior more systematically, we perform detailed temperature-dependent measurements of the energy gap $\Delta(T)$ using ARPES. In Fig. 2, we show a quantitative comparison between $\Delta(T)$ and $R_{s}(T)$ measurements on the same sample shown in Fig. 1. In Fig. 2a, we plot over 100 energy distribution curves (EDCs) symmetrized about $E_{\mathrm{F}}$ from 12 K to 94 K, measured at $k_{\mathrm{F}}$ of the electron pocket, where false color represents the intensity of the EDCs.  In Fig. 2b, we plot select EDCs extracted from the temperature series in Fig. 2a.  Fig. 2c tracks $\Delta$ as a function of temperature, defined as half the separation between quasiparticle peaks of the symmetrized EDCs from Figs. 2a and 2b, as well as the evolution of the spectral gap depth $\delta_{SW}$, defined as the difference between the coherence peak amplitude normalized to unity and the corresponding spectral weight at $E_F$.  In Fig. 2d we show $R_{s}(T)$, as well as its derivative $dR_{s}/dT$.  As the superconducting transition is broad, we define three characteristic temperatures to describe the shape of the transition: \Tzero, where the resistance reaches 0.1$\%$ of $R_{s}(70 \mathrm{K})$; \Tonsetns, the intersection between the extrapolated normal-state sheet resistance and a linear fit to the superconducting transition region; and \Tstarns, where $R_{s}(T)$ exhibits an inflection point in the paraconducting state (as determined by a local minimum in $dR_{s}/dT$).  For the sample shown in Fig. 2, \Tzero = 29$\pm0.2$ K, while \Tonset = 44$\pm3$ K, and \Tstar = 72$\pm4$ K.  Deep within the superconducting state (T $<$ \Tzerons), a clear superconducting gap ($\Delta = 12.8\pm1$ meV) and sharp Bogoliubov quasiparticle peaks are observed in the ARPES spectra.  In the broad transition region where \Tzero $ < T < $ \Tonsetns, the strength of the quasiparticle peak is gradually suppressed as temperature increases, accompanied by a rapid filling of spectral weight within the gap (Fig. 2, a and c), despite the energy separation between the peaks remaining largely constant.  Upon increasing the temperature further (\Tonset $ < T < $ \Tstarns), the energy gap continues to fill in at a more gradual rate, until eventually $\Delta$ is no longer discernible above \Tgap = 73$\pm$5 K, a temperature that corresponds closely to \Tstarns.  

This behavior is in stark contrast to what is observed in bulk conventional superconductors, where the resistivity drops abruptly to zero at the same temperature at which the superconducting gap opens (i.e, \Tzero $\approx$ \Tonset $\equiv$ \Tgapns). The most notable exception to this is underdoped cuprates, where the pseudogap at the $d$-wave antinode measured by numerous techniques including ARPES also opens at significantly higher temperatures than the bulk \TCnospace~\cite{Loeser1996}. In contrast, in bulk Fe-based superconductors, it has been widely shown that \Tonset and \Tgap match closely~\cite{Ding_2008,stewart2011}, including in electron-doped bulk FeSe-based compounds such as $A_x$Fe$_2$Se$_2$ ($A$ = K, Cs) and (Li$_{1-x}$Fe$_x$)OHFeSe~\cite{Zhang2011,Zhao2016}.  Thus the observed discrepancy in \FeSeSTO cannot be solely attributed to the unconventional nature of Fe-based superconductivity. 

On the other hand, such behavior is expected in 2D superconductors which can exhibit a broad Berezinskii-Kosterlitz-Thouless~(BKT) transition~\cite{Kosterlitz1973}, where vortex-antivortex fluctuations prevent long range phase coherence at temperatures well above where a zero resistance state is finally achieved (\Tbktns). BKT transitions have been extensively studied in disordered 2D superconductors as well as more recently in atomically thin crystalline superconductors or interfaces such as LaAlO$_3$/SrTiO$_3$~\cite{LAOSTO_2007} and twisted bilayer graphene~\cite{Cao2018}. Probes such as ARPES or tunneling spectroscopy detect the initial formation of pairs, but are not sensitive to their phase coherence, so a spectroscopic gap can be found to open at temperatures well above a broad resistive transition (\Tgap $>$ \Tbktns). Recently, combined tunneling and transport measurements of disordered ultrathin films of the BCS superconductors TiN~\cite{Sacape2010} and NbN~\cite{mondal2011} have verified such a picture.  

To quantitatively investigate the possibility of BKT phase fluctuations in \FeSeSTOns, we show $V$($I$) characteristics from the \FeSeSTO films in Fig. 3 on a log-log scale, measured from 24 to 37 K. The slopes of the curves in Fig. 3a indicate the power law exponent $\alpha$ at low voltages for $V(I) \propto I^{\alpha}$ (Fig. 3b).  As expected for a BKT-like transition, the values of $\alpha$ are highly temperature dependent, crossing $\alpha = 3$ at \Tbkt = $27.1\pm0.5$ K.  A plot of $[d(\ln R)/dT]^{2/3}$ (Fig. 3c) matching the Halperin-Nelson form of $R_{s}(T)$~\cite{Halperin1979} yields a value of \Tbkt $= 27.2\pm0.5$ K, in agreement with \Tbkt extracted from the critical exponent analysis.  Signatures of a BKT transition have also been reported in \emph{ex situ} measurements of capped \FeSeSTO thin films, albeit with lower values of \Tonset and \Tbktns~\cite{Zhao_2016}.   

Since \Tzero in 2D superconductors can be strongly influenced by disorder, we have systematically investigated a large number of samples with varying degrees of disorder, using the extrapolated residual sheet resistivity $R_0$ as a metric, and controlled primarily through the post-growth annealing process~\cite{Huang2016}.  A comparison with ARPES data shows close correspondence between $R_0$ and increased quasiparticle broadening, consistent with sample-to-sample variation in the disorder strength (Fig. S6).  In Figure 4a, we show $R_s$($T$) for a selection of single-layer \FeSeSTO films, which clearly demonstrates the obvious dependence of \Tzero and \Tonset on $R_0$. As shown in Figure 4c, which summarizes all samples measured in this study, \Tzero decreases linearly with increasing $R_0$, approaching 40 K in the clean limit. The crossover from a superconducting to insulating regime occurs around $R_0$ $\approx$ 7.2 k$\Omega$, close to the quantum of resistance for pairs, $R_Q = h/(2e)^2$, as would be expected for a 2D superconductor limited by phase fluctuations~\cite{PhysRevLett.64.587}. The importance of disorder on 2D phase fluctuations naturally explains the wide variation in \Tzero and \Tonset values \cite{Zhang2014,Yang2019,Zhao_2016,Zhao_SciAdv_2018,Wang_2017} previously reported in the literature from capped films (Fig 4b).  The highest values of \Tonset $\approx$ 45 K reported here on pristine films are slightly higher than the maximum \Tonset observed in capped films from the literature ($\approx$ 40 K), and are inconsistent with the singular report of \Tc $> 100$ K by Ge \emph{et al.} \cite{Ge2014}. 

In contrast to \Tzero, both \Tgap and \Tstar show relatively little dependence on disorder (Fig. 4c), with the values of \Tgap reported here generally consistent with the values extracted from the literature using the same analysis method for our own data (Fig. 4b, grey symbols)~\cite{Liu_origin_2012,Song2019,Tan2013,He2014,Peng2014,Lee2014,Peng2015}.  The close correspondence of \Tgap and \Tstar strongly suggests that the beginning of the resistive transition at \Tstar is directly related to the appearance of Cooper pairs below \Tgapns.  This incoherent Cooper pairing persists within a high temperature pseudogap regime (\Tonset $<$ T $<$ \Tgap) well above the temperature range where 2D BKT-like phase fluctuations are clearly observed (T $<$ 40 K).  

Taken together, these measurements present, for the first time, a self-consistent picture for the previously mysterious superconducting behavior of \FeSeSTOns.  At low temperatures (T $<$ \Tzero), the influence of phase fluctuations is minimal, resulting in sharp Bogoliubov quasiparticle peaks and a zero resistance state. As the temperature is increased, the zero resistance state is destroyed by a BKT-like vortex unbinding transition, at a temperature dependent on the level of disorder, while spectral weight begins to fill within the gap. Since \Tzero should asymptote to $T_{c}$ in the clean limit for a 2D superconductor~\cite{beasley1979}, this suggests that the maximum intrinsic \Tc of \FeSeSTO is $\approx$ 40 K, when accounting for disorder and phase fluctuations (Figure 4c), comparable to the highest values of \Tc for bulk electron-doped FeSe-based compounds (Figure 4b) such as (Li$_{1-x}$Fe$_x$)OHFeSe~\cite{Zhang2011}, but well short of the 60-70 K \Tc previously interpreted from spectroscopic results alone.

Finally, we speculate on the high-temperature pseudogap regime for \FeSeSTO (\Tonset $< T <$ \Tgap $\approx$ 60-70 K), when compared to bulk electron-doped FeSe-based materials which do not exhibit a pseudogap and show \Tgap $\approx$ \Tc $\approx$ 40 K.  One possibility is that Gaussian fluctuations above \Tc account for the behavior observed in the high-temperature pseudogap regime.  However this scenario is contradicted by the observed behavior of $\Delta(T)$, which shows no evidence of closing near 40 K.  Instead, this suggests that the microscopic, mean-field pairing temperature of \FeSeSTO is intrinsically higher than that of bulk FeSe-based compounds, even if the ultimate maximum \Tc $\approx$ 40 K for \FeSeSTO is comparable to those of bulk compounds. Much speculation has focused on the possible influence of interfacial electron-phonon coupling from the SrTiO$_3$ substrate in enhancing the \TCnospace~\cite{Lee2014,Song2019}. Alternatively, recent work on the highly two-dimensional bulk compound (TBA$^{+}$)FeSe, where the distance between FeSe layers is expanded to 15.5 \AA\ by  intercalation of ion tetrabutyl ammonium organic molecules (compared to 5.5 \AA\ for bulk FeSe or 9.32 \AA\ for (Li$_{1-x}$Fe$_x$)OHFeSe), has also reported evidence of incoherent pre-formed pairing up to 60 K, comparable to \FeSeSTOns, but in the absence of any substrate~\cite{kang2019preformed}. This suggests that the increased two-dimensional nature of the electronic or crystal structure could potentially be the origin of the enhanced mean-field pairing temperature \Tgap in \FeSeSTOns. While it is empirically known that a two-dimensional electronic structure appears to be a key ingredient for unconventional high-temperature superconductivity (e.g. cuprates, Fe-based superconductors, nickelates), most Fe-based superconductors exhibit some degree of three-dimensionality in their electronic structure, as evidenced by $k_z$ dispersion in ARPES~\cite{Liu2009}, as well as their resistivity anisotropy $\rho_c / \rho_{ab}$ being in the range of 2-3 for the 11 and 111 families or up to $10^{2}$ for the 122 compounds~\cite{PhysRevB.87.134512,song2010,PhysRevB.78.224512}. This is in contrast to their more two-dimensional, higher \Tc cuprate analogues such as Bi$_{2}$Sr$_{2}$CaCu$_{2}$O$_{8 + \delta}$, YBa$_{2}$Cu$_{3}$O$_{7-\delta}$, or La$_{2-x}$Sr$_x$CuO$_4$, where $\rho_c / \rho_{ab}$ is much larger, in the range of 10$^3$-10$^6$~\cite{PhysRevB.65.214535}. By pushing Fe-based superconductors closer to the idealized two-dimensional limit as in (TBA+)FeSe ($\rho_c / \rho_{ab} \approx $ 10$^5$) or in the ultimate case of single-layer \FeSeSTOns, it is possible that the strength of the microscopic pairing is increased, but at the cost of 2D phase fluctuations and enhanced sensitivity to disorder which limit \Tzerons.    

\bibliography{FeSe_SC}

\section*{Acknowledgments}
We would like to thank P.B. Littlewood for helpful discussions. This work was supported through the Air Force Office of Scientific Research Grant No. FA9550-15-1-0474, and the National Science Foundation Platform for the Accelerated Realization, Analysis, and Discovery of Interface Materials (PARADIM) under Cooperative Agreement No. DMR-1539918, NSF DMR-1709255. This research is funded in part by the Gordon and Betty Moore Foundation's EPiQS Initiative through Grant No. GBMF3850 to Cornell University. B.D.F. and J.N.N. acknowledge support from the NSF Graduate Research Fellowship under Grant No. DGE-1650441. P.M. acknowledges support from the Indo US Science and Technology Forum (IUSSTF).  This work made use of the Cornell Center for Materials Research (CCMR) Shared Facilities, which are supported through the NSF MRSEC Program (No. DMR-1719875). Substrate preparation was performed in part at the Cornell NanoScale Facility, a member of the National Nanotechnology Coordinated Infrastructure (NNCI), which is supported by the NSF (Grant No. ECCS-1542081). 


\section{Methods}

\subsection{Film synthesis}

Single-layer \FeSeSTO films are synthesized on SrTiO$_3$ (001) substrates using a chalcogenide molecular beam epitaxy (MBE) system as reported previously~\cite{Lee2014}.  Se (99.999$\%$ purity) and Fe (99.995$\%$ purity) are co-evaporated at a nominal flux ratio of 5:1 and at a nominal growth rate of 1.8-2 unit cells (uc) per minute, with source fluxes calibrated by a quartz crystal monitor (QCM), and film crystallinity monitored in real time using reflection high energy electron diffraction (RHEED).  To enable reliable resistivity measurements of the FeSe monolayer, we utilized undoped insulating SrTiO$_3$ for all films presented in this work.  After growth films are progressively annealed until optimal superconducting properties are achieved, followed by deposition of 20 nm thick Au electrodes at the sample corners using a shadow mask to provide reliable 4-point electrical contact.  

\subsection{\insitu electrical transport measurement}

\textit{In situ} resistivity measurements are preformed using a custom-built UHV 4-point transport probe with a base temperature of 5.2~K and a base pressure of \smash{7 $\times$ 10$^{-11}$} Torr.  Contact is applied directly to the film using a set of Au-plated spring-loaded probes in a Van der Pauw geometry, with a nominal instrumental contact spacing of 7~mm.  Resistance measurements were taken using a Keithley 6221/2182A current source/voltmeter combination in delta mode with a typical applied current of 1-10~uA.  Sheet resistance values were calculated from the Van der Pauw equation such that:
\begin{equation}
R_{s}=\frac{1.34 \pi}{\ln (2)} \frac{V_{24}}{I_{13}},
\end{equation}
where \( \frac{\pi}{\ln (2)} \) is the Van der Pauw factor and 1.34 is an additional factor to account for the finite contact dimensions based on the known dimensions of the Au electrodes~\cite{chwang1974}.  A subset of the films analyzed in Figure 4 were measured without gold electrodes present; in this case we instead use a correction factor of 1.1, based on finite-element analysis of the Van der Pauw correction factor for our known probe geometry.

\subsection{ARPES Measurements}

ARPES measurements were taken with a VG Scienta R4000 electron analyzer equipped with a VUV5000 Helium discharge lamp using He-I photons at 21.2~eV.  The base pressure in the ARPES system is \smash{5 $\times$ 10$^{-11}$}~Torr.  Energy resolution was nominally set at 12~meV for mapping and 9~meV for gap measurements.  To avoid sample charging during ARPES measurement the film is grounded using a retractable contact pin built onto the sample manipulator.  For gap measurements, the Fermi level is referenced to the measured Fermi edge of the Au electrodes.  

\newpage
\null
\vfill

\begin{figure*}
\begin{center}
\includegraphics[width=7in]{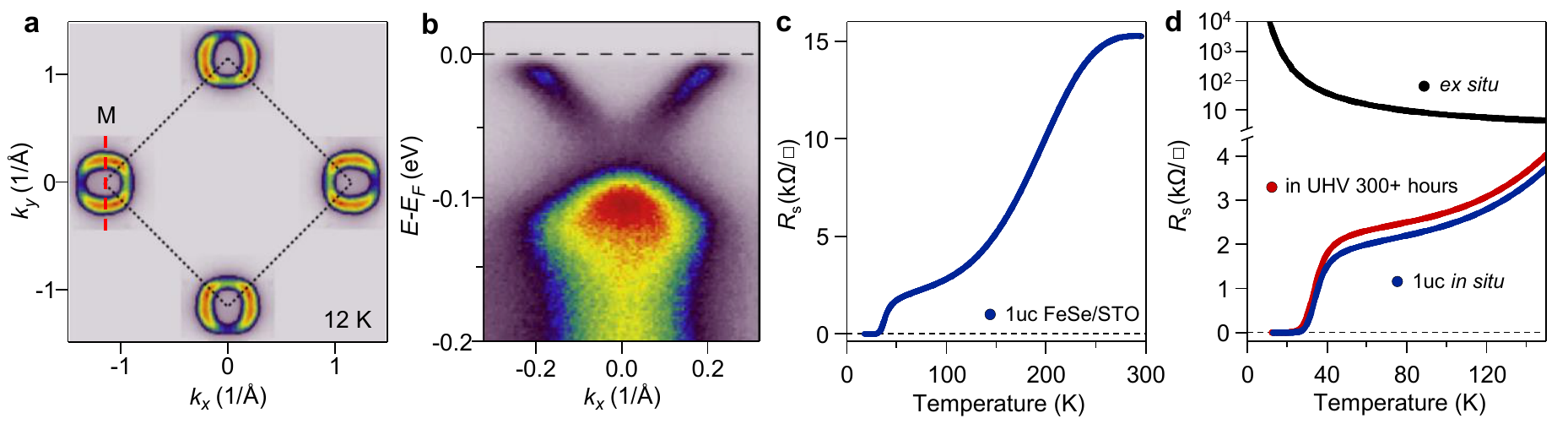}
\end{center}
\caption{
\textbf{	Combined ARPES and \textit{in situ} electrical resistivity measurement of single-layer \FeSeSTOns.} Measurements in panels A-C are conducted on the same sample.  \textbf{(a)} Fermi surface intensity map for an as-grown 1 uc \FeSeSTO sample held at 12 K, integrated over $\pm$5 meV of $E_F$.  The black dashed line indicates the boundary of the 2-Fe Brillouin zone.  The electron pockets comprise 5.5\% of the Brillouin zone area.  \textbf{(b)} Photoemission intensity at M (dashed red line in A) taken at 12 K. \textbf{(c)} Temperature-dependent sheet resistance for 1 uc \FeSeSTOns.  \textbf{(d)} Stability of the superconducting state for a separate sample after growth (blue), long-term UHV storage (red), and after momentary exposure to inert gas atmosphere (black).
}
\label{fig:Fig1}
\end{figure*}

\vfill

\begin{figure*}
\begin{center}
\includegraphics[width=6.5in]{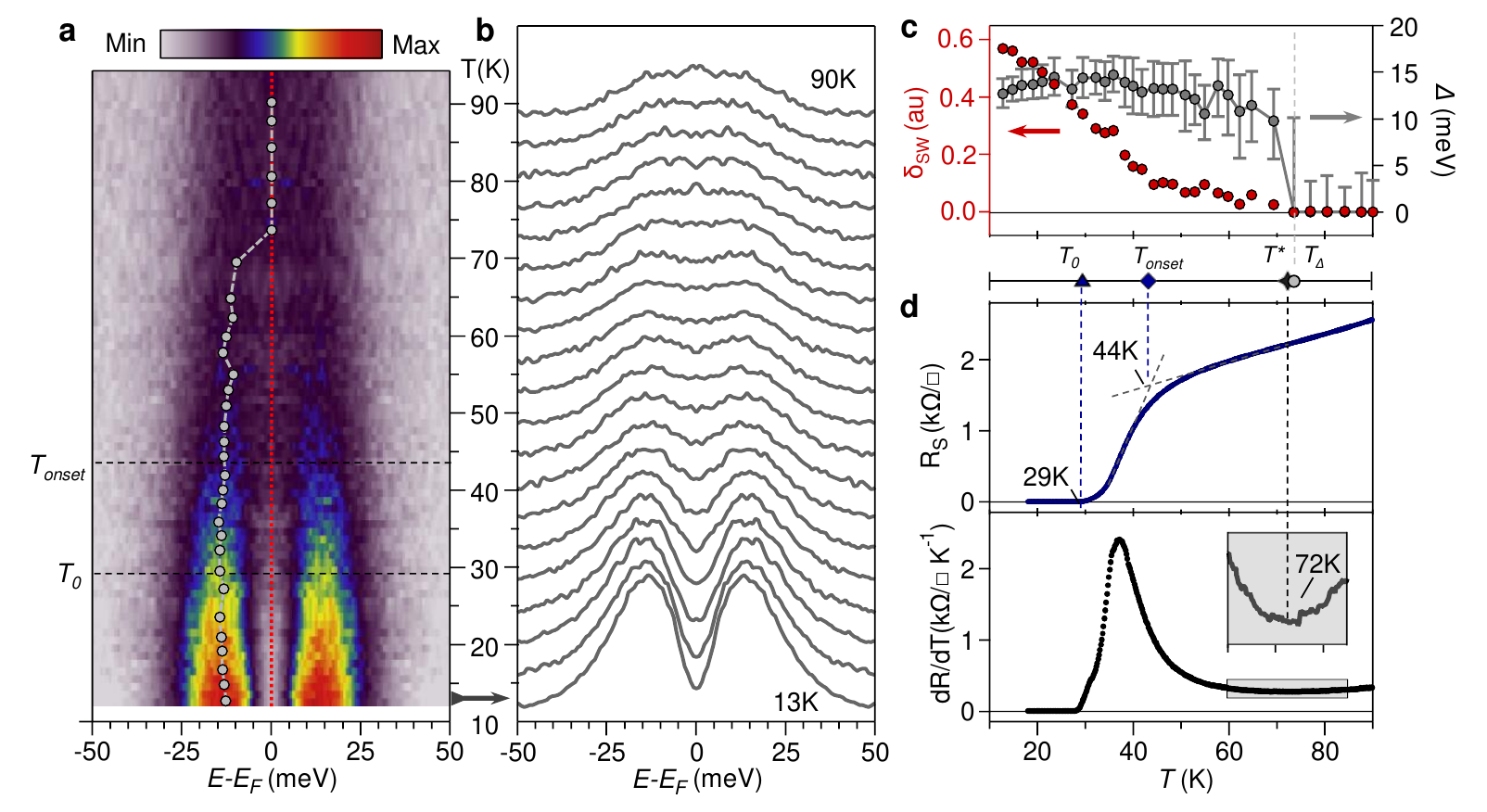}
\end{center}
\caption{
\textbf{Comparison of temperature-dependent gap closing behavior and corresponding resistive superconducting transition.} \textbf{(a)} Symmetrized EDC's taken at $k_F$ from over 100 individual spectra collected between 12-94 K.  The color scale indicates the EDC intensity, with increasing temperature along the vertical axis.  The grey circles track the quasiparticle peak position as a function of temperature.  \textbf{(b)} Selection of symmetrized EDC’s from the data in \textbf{(a)}.  The temperature (vertical) axis is matched to that of \textbf{(a)}.   \textbf{(c)} Extracted energy gap $\Delta$ (right axis) as a function of temperature from the data in (a). $\Delta$ is defined as half of the separation between the EDC peak positions indicated in (a).  The red symbols (left axis) track the suppression spectral weight $\delta_{SW}$ at $E_{\mathrm{F}}$ (dotted line in a) relative to 90 K.  \textbf{(d)} Sheet resistance (blue, top) and d$R$/d$T$ (black, bottom) as a function of temperature measured \emph{in situ} for the same sample presented in (a) and (b).
}
\label{fig:Fig2}
\end{figure*}

\begin{figure*}
\begin{center}
\includegraphics[width=6.5in]{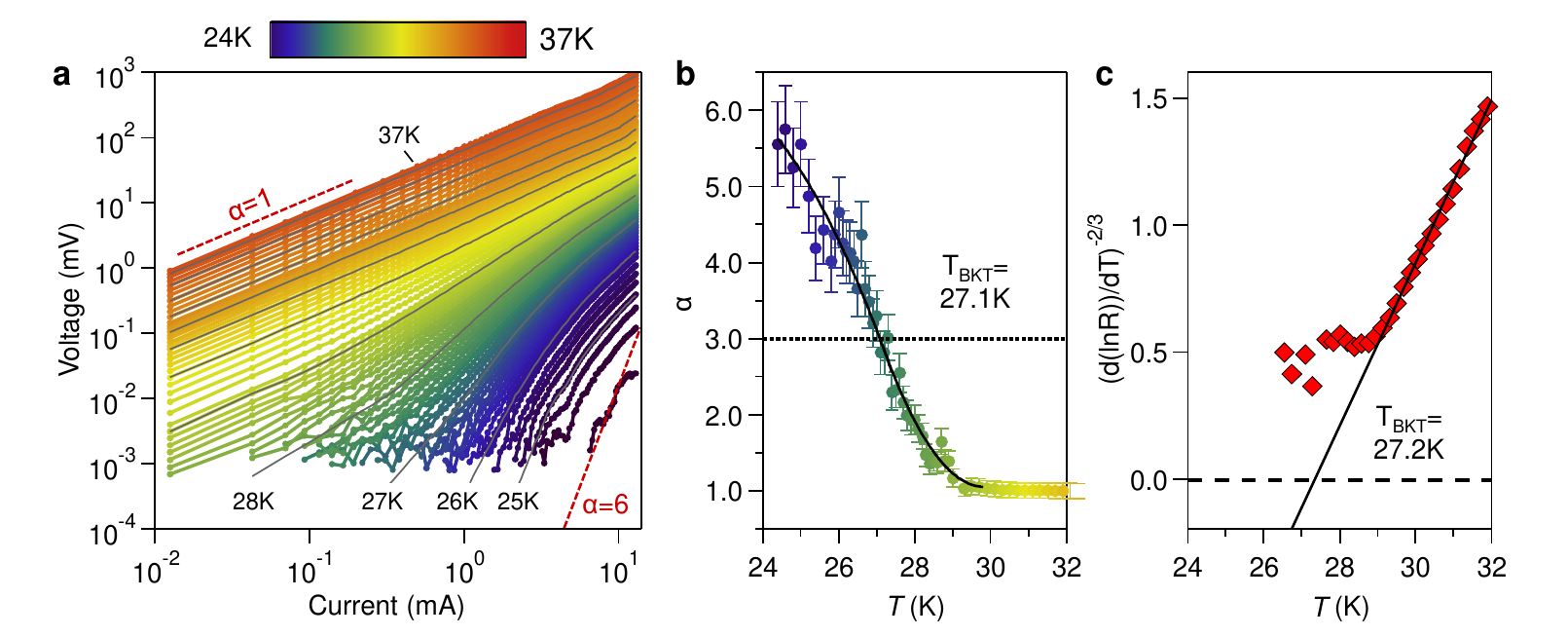}
\end{center}
\caption{
\textbf{ \insitu $V$($I$) characteristics and $R$($T$) analysis.} \textbf{(a)} Current-voltage characteristics for 1 uc thick \FeSeSTOns, shown on log-log scale.  Color indicates the temperature for a given $V-I$ curve, where the grey lines indicate temperature spacings of 1 K.  Red dashed lines indicate power-law behavior for $V\propto I^{\alpha}$. \textbf{(b)} Extracted temperature dependence of $\alpha$ from power-law fits to the data in \textbf{(a)}.  \textbf{(c)} $[d(lnR)/dT]^{2/3}$ plotted vs $T$.  The solid black line indicates behavior consistent with $T_{BKT}$=27.2 K.    
}
\label{fig:Fig3}
\end{figure*}

\begin{figure*}
\begin{center}
\includegraphics[width=6.5in]{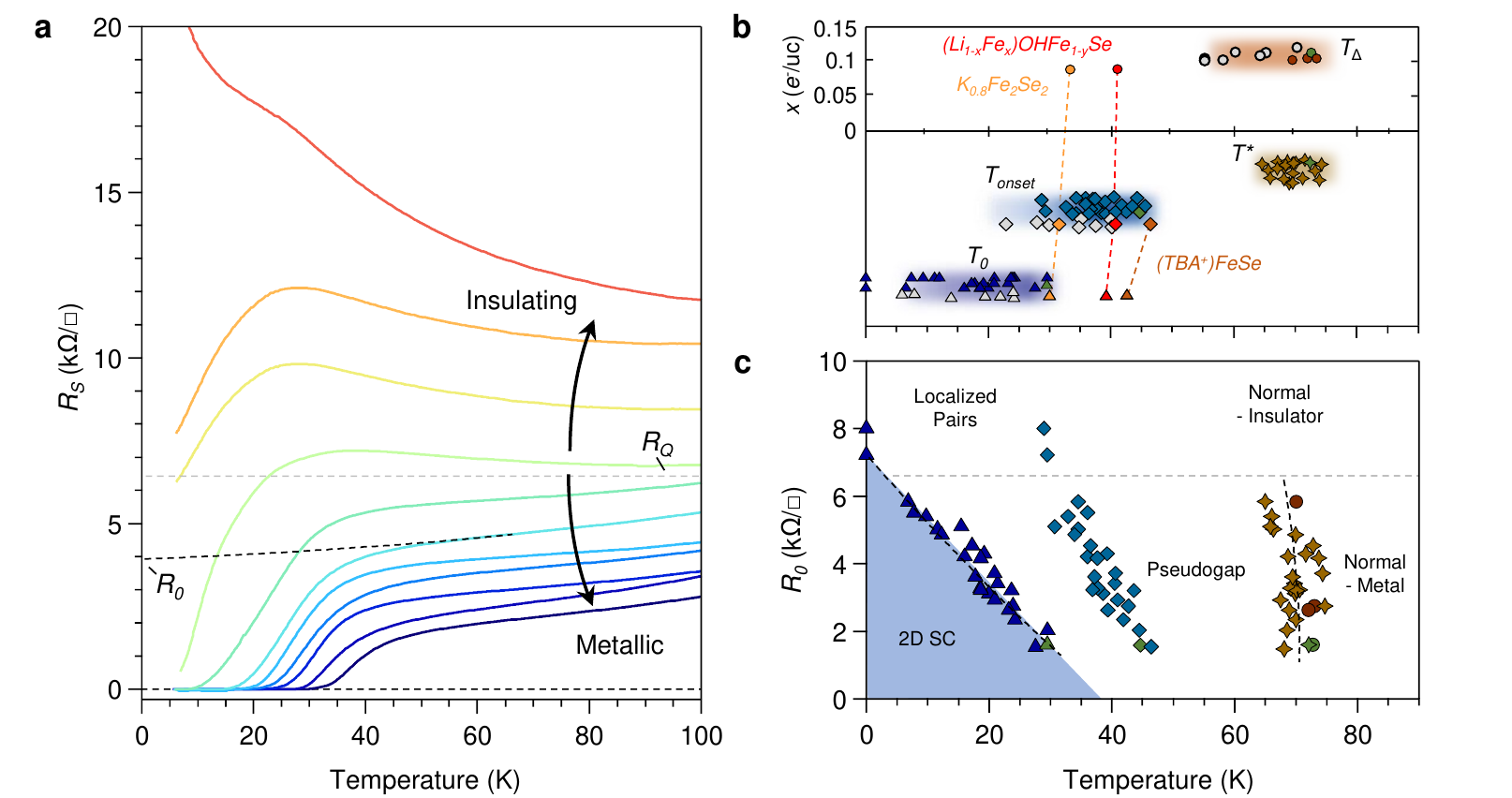}
\end{center}
\caption{
\textbf{Disorder driven superconductor-insulator transition in single-layer \FeSeSTOns.}  (\textbf{a}) \Rsheet for a selection of single-layer \FeSeSTO films.  The residual resistance $R_0$ for each curve is determined by an extrapolation of the normal state data above \Tstar (black dashed line).  (\textbf{b}) Extracted characteristic temperatures of single-layer \FeSeSTOns.  Marked symbols indicate \Tzero (triangles), \Tonset(diamonds), \Tstar(stars), and \Tgap(circles) vs x (Luttinger volume) as measured \insitu by this study (solid colored symbols) and from the literature (gray filled symbols).  \Tzerons, \Tonsetns, and \Tstar values are offset along the Y axis for clarity.  Literature data for \Tzero and \Tonset are taken from capped single-layer \FeSeSTO films. Green symbols indicate values for the sample presented in Figs 1-3.  ARPES gap values and doping levels from the literature are reported from ~\cite{Liu_origin_2012,Song2019,Tan2013,He2014,Peng2014,Lee2014,Peng2015}.  $Ex$ $situ$ transport data from \cite{Zhang2014,Yang2019,Zhao_2016,Zhao_SciAdv_2018,Wang_2017}.  \Tzerons, \Tonsetns, and \Tgap values are also plotted for bulk compounds K$_{0.8}$Fe$_2$Se$_2$ (orange, solid), (Li$_{1-x}$Fe$_x$)OHFeSe (red, solid), and (TBA+)FeSe (brown, solid).  (\textbf{c}) $R_0$ vs \Tzerons, \Tonsetns, \Tstarns, and \Tgap for films measured \insitu for this work.
}
\label{fig:Fig4}
\end{figure*}

\end{document}